\documentclass[aps,twocolumn,longbibliography,english,prx]{revtex4-1}
\usepackage[colorlinks=true,urlcolor=blue,citecolor=blue,linkcolor=blue]{hyperref}
\usepackage[T1]{fontenc}
\usepackage{amssymb}
\usepackage{graphicx}
\usepackage{amsmath,color}
\usepackage{mathrsfs}
\usepackage{float}
\usepackage[normalem]{ulem}
\usepackage{indentfirst}
\usepackage{txfonts}
\usepackage{algpseudocode}
\usepackage{algorithm}
\usepackage{booktabs}

\emergencystretch=\maxdimen
\makeatletter

\def\journal #1, #2, #3, 1#4#5#6{{\sl #1~}{\bf #2}, #3 (1#4#5#6) }

\newcommand{\BAS}{Bars-and-Stripes }

\newcommand{\Eq}[1]{Eq.~(\ref{#1})}

\newcommand{\Ref}[1]{Ref.~\onlinecite{#1}}
\newcommand{\Sec}[1]{Sec.~\ref{#1}}

\newcommand{\material}[1]{\iffalse[{\bf  \color{cyan}{Material: #1}}]\fi}

\usepackage{amsmath, amssymb}
\DeclareMathOperator{\E}{\mathbb{E}}

\makeatother

\usepackage{babel}

\begin{document}
\title{Learning and Inference on Generative Adversarial Quantum Circuits}

\author{Jinfeng Zeng$^{1,2}$}
\author{Yufeng Wu$^{3}$}
\author{Jin-Guo Liu$^{1}$}
\email{cacate0129@iphy.ac.cn}
\author{Lei Wang$^{1,4}$}
\email{wanglei@iphy.ac.cn}
\author{Jiangping Hu$^{1,5,6}$}

\affiliation{$^{1}$ Institute of Physics, Chinese Academy of Sciences, Beijing 100190, China}
\affiliation{$^{2}$ University of Chinese Academy of Sciences, Beijing 100049, China}
\affiliation{$^{3}$ College of Physics, Jilin University, Changchun 130012, China}
\affiliation{$^{4}$ CAS Center for Excellence in Topological Quantum Computation, University of Chinese Academy of Sciences, Beijing 100190, China}
\affiliation{$^{5}$ Kavli Institute of Theoretical Sciences, University of Chinese Academy of Sciences, Beijing, 100190, China}
\affiliation{$^{6}$ Collaborative Innovation Center of Quantum Matter, Beijing, China}

\begin{abstract}
Quantum mechanics is inherently probabilistic in light of Born's rule. Using quantum circuits as probabilistic generative models for classical data exploits their superior expressibility and efficient direct sampling ability. However, training of quantum circuits can be more challenging compared to classical neural networks due to lack of efficient differentiable learning algorithm.
We devise an adversarial quantum-classical hybrid training scheme via coupling a quantum circuit generator and a classical neural network discriminator together. After training, the quantum circuit generative model can infer missing data with quadratic speed up via amplitude amplification. We numerically simulate the learning and inference of generative adversarial quantum circuit using the prototypical \BAS dataset.
Generative adversarial quantum circuits is a fresh approach to machine learning which may enjoy the practically useful quantum advantage on near-term quantum devices.
\end{abstract}

\maketitle

\section{Introduction}
Probabilistic generative modeling~\cite{Goodfellow2016, Goodfellow2016c} is a major direction of deep learning research~\cite{LeCun2015} towards the goal of artificial general intelligence. Generative models capture all patterns that are present in the data by modeling the full joint probability distribution. Moreover, they can even generate new samples according to the learned probability distribution. Generative models find wide applications in complex tasks beyond classification and regression, such as speech synthesis~\cite{Van2016} and image-to-image translation~\cite{Zhu2017}. A typical application of generative models is the inference, in which the model infers the missing data conditioned on partial observations. For example, inpainting missing or deteriorated parts of images according to the learned conditional probability~\cite{Pathak2016}.

Generative adversarial network (GAN)~\cite{Goodfellow2014} is one of the most popular generative models. GANs cast the generative modeling as a two-player game: a generator produces synthetic data to mimic the distribution of the training data and a discriminator differentiates the synthetic data from the training data. The generator acts as a simulator, which implicitly defines a learned model probability distribution. Compared to generative models with explicit likelihood such as the Boltzmann Machines~\cite{Ackley1985}, autoregressive models~\cite{Graves2013, Van2016PixelRNN, Van2016PixelCNN}, normalizing flows~\cite{Dinh2014, Dinh2016, Glow2018}, and the variational autoencoders~\cite{Kingma2013,Rezende2014}, GANs have more flexibility in the network structure and hence stronger expressibility. A common thread of these classical generative models~\cite{Goodfellow2014, Ackley1985, Graves2013, Van2016PixelRNN, Van2016PixelCNN, Dinh2014, Dinh2016, Glow2018, Kingma2013,Rezende2014} is that they all express or transform probability distributions using neural networks.

Quantum circuit Born machine (QCBM)~\cite{Benedetti2018,Liu2018} leverages the probabilistic nature of quantum mechanics for generative modeling. QCBM represents classical probability distribution using quantum pure states according to the Born's rule~\cite{Cheng2017,Han2017}. Thus, in contrast to the classical neural networks, the QCBM expresses probability by transforming wavefunction via a sequence of unitary gates. Afterwards, one can directly obtain samples by projective measurement of the QCBM on the computational basis. Since the probabilistic output of a general quantum circuit cannot be simulated efficiently by classical computation, the QCBM can be regarded as practical quantum machine learning application of the quantum supremacy~\cite{Boixo2018}.
However, similar to the classical GANs, the model probability density of the QCBM is implicit since one does not have access to the wave-function of the actual quantum circuit. \Ref{Liu2018} developed a differentiable learning technique of QCBM using a hybrid quantum-classical methods by minimizing the kernel maximum mean discrepancy (MMD) loss~\cite{Gretton2007,Gretton2012} in the line of the generative moment matching networks~\cite{Li2015,Dziugaite2015}.

In this paper, we develop an adversarial training scheme for QCBM, and perform quantum inference on the trained quantum circuit. The training approach is an instance of the quantum-classical hybrid algorithms~\cite{Peruzzo2014, McClean2016VQE, Kandala2017, Farhi2014, Farhi2017, Otterbach2017, Mitarai2018, Benedetti2018}, which involves a game between a quantum circuit generator and a classical neural network discriminator. 
The quantum generator produces a probability distribution in the computational basis to mimic the one of the classical dataset. The classical discriminator differentiates whether its input is produced by the quantum circuit or from the training dataset. We devise an unbiased gradient estimator for the generative adversarial loss and perform differentiable learning of the discriminator and the generator. An advantage compared to the previous work~\cite{Liu2018} is that the computational cost of the GANs loss is linear in terms of the number of samples other than the quadratic scaling MMD loss. Also, training of GANs requires smaller batch size. 
After training, one can perform accelerated quantum inference~\cite{Low2014} on the quantum circuit to infer the value of unobserved data conditioned on partial observations.

There have been some recent works \cite{Lloyd2018, Demers2018QGAN, Benedetti2018Gan, Hu2018} on generalization the GANs to quantum domain. \Ref{Lloyd2018} describes three theoretical protocols of quantum generative adversarial learning, in which the dataset, discriminator, and the generator could be classical or quantum. While the authors of ~\cite{Demers2018QGAN, Benedetti2018Gan} consider quantum data and quantum circuits for both the discriminator and the generator. \Ref{Hu2018} demonstrated experimental realization of quantum GAN using superconducting quantum qubit. Our paper is different  since we focus on modeling of classical dataset in real-world machine learning applications using quantum circuits in a scalable way. And we employ a classical neural network as the discriminator, which leads to a hybrid quantum-classical optimization scheme. The reasoning of having a quantum generator and a classical discriminator is that one believes that the complexity of generative tasks is higher than discriminative tasks.

The organization of this paper is as follows. In \Sec{sec-model} we first present the framework of generative adversarial quantum circuit and its differential learning scheme. Next, we describe quantum inference on a trained quantum circuit using amplitude amplification. Then, in \Sec{sec-result} we demonstrate the adversarial training and image inpainting application using the \BAS dataset. Finally, we outlook for future research directions in \Sec{sec-discussion}.

\section{Theoretical Framework} \label{sec-model}


In this section, we first introduce the general structure of generative adversarial quantum circuit, and then describe the learning and inference on the quantum circuit.

\subsection{Generative Adversarial Quantum Circuits}\label{model}

\begin{figure}[tb]
\centerline{\includegraphics[width=0.48\textwidth]{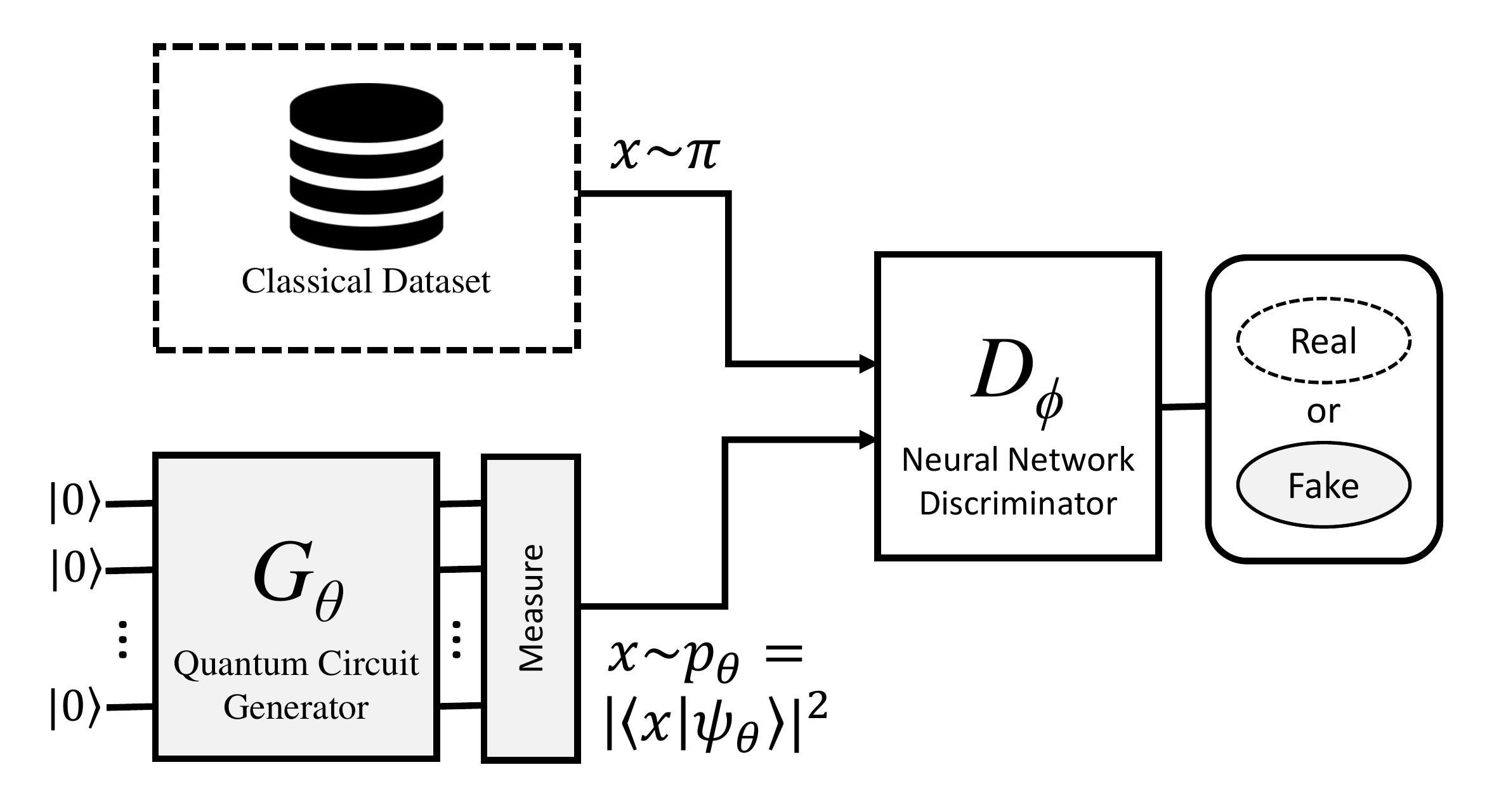}}
\caption{Schematic illustration of the generative adversarial training of quantum circuit. The generator $G_{\boldsymbol{\theta}}$ is a quantum circuit which evolves the input state $|z\rangle=|0\rangle^{\otimes N}$ to a final state $|\psi_{\boldsymbol{\theta}}\rangle$ via a sequence of parametrized unitary gates. Projective measurement of the final state on the computational basis yield synthetic bitstring samples. The discriminator $D_{\boldsymbol{\phi}}$ is a classical neural network, which differentiates between the real samples in the training dataset and the synthetic samples generated by the quantum circuit.
}
\label{gan}
\end{figure}
As shown in Fig.~\ref{gan}, the generative adversarial quantum circuit consists of two components, a generator G and a discriminator D. The major difference compared to the classical GAN~\cite{Goodfellow2014, Goodfellow2016} is that we use a quantum circuit as the generator of classical data. The quantum circuit takes the initial state $|z\rangle = |0\rangle^{\otimes N}$ as an input and evolves it to a final state $|\psi_{\boldsymbol{\theta}}\rangle = U_{\boldsymbol{\theta}} |z\rangle$ via a sequence of unitary gates. Measuring of the output state on the computational basis produces samples of bitstrings $x\thicksim p_{\boldsymbol{\theta}}(x) = |\langle x | \psi_{\boldsymbol{\theta}} \rangle|^{2}$.  The discriminator D is a classical neural network which takes either real samples from the training dataset or the synthetic samples from the quantum circuit as inputs. 
The discriminator D performs a binary classification on whether its input $x$ is from classical training dataset [$D_{\boldsymbol{\phi}}(x)=1$] or generated by quantum circuit  [$D_{\boldsymbol{\phi}}(x)=0$]. While the goal of G is to generate synthetic samples to fool the discriminator, which means the G strives to produce samples $x$ with $D_{\boldsymbol{\phi}}(x)=1$. The generator and the discriminator, implemented as a QCBM and a classical neural network, play against each other to achieve their individual goals. This adversarial game defines a  minimax problem \Eq{minmax} with parametrized quantum circuit $G_{\boldsymbol{\theta}}$  and classical neural network  $D_{\boldsymbol{\phi}}$, and objective function reads
\begin{eqnarray}
\min_{G_{\boldsymbol{\theta}}} \max_{D_{\boldsymbol{\phi}}} \E_{x\thicksim\pi}[ \ln D_{\boldsymbol{\phi}}(x)] + \E_{x \thicksim p_{\boldsymbol{\theta}}}[\ln (1 -  D_{\boldsymbol{\phi}}(x))]\label{minmax}.
\end{eqnarray}
Assuming that both the quantum circuit G and the neural network D have sufficient capacity, the minimax game has a Nash equilibrium~\cite{Goodfellow2014}. The equilibrium is an optimal solution where the D cannot discriminate whether its input is real or fake and outputs $D_{\boldsymbol{\phi}}={1}/{2}$ for all inputs. And the quantum circuit G creates samples in accordance with the distribution of the training dataset.

For the generator, we use quantum circuit architecture similar to previous studies~\cite{Liu2018, Kandala2017}.
The quantum circuit contains the single qubit rotation layers and the two qubits entangler layers.
The single qubit rotation layers are formulate as $U_{\boldsymbol{\theta}}=\{U_{\theta_{\alpha}^{l}}\}$, where the layer index $l$ runs from $0$ to $d$ and $\alpha$ is a combination of qubit index $i$ and rotation gate index $j$. In the layer $l$, the rotation gate has the form
$U_{\theta_{\alpha}^l} = U_{\theta_{i,j}^l} = R_z(\theta_{i,1}^l)R_x(\theta_{i,2}^l)R_z(\theta_{i,3}^l)$ with $R_m(\theta) = \exp \left(\frac{-i\theta\sigma_m}{2}\right)$ , where the $\sigma_m$$(m={x,y,z})$ are Pauli matrices and $i$ runs from 0 to $N-1$, with $N$ the number of qubits. We use the CNOT gates with no learnable parameters as the entangle layers to induce correlations between qubits. We then stack alternately the single qubit rotation layer and the entangler layer to construct the quantum circuit and let the last layer to be the single qubit rotation layer. The total number of learnable parameters in this quantum circuit is $(3d+1)N$, where $d$ is the maximum depth of the circuit.

For the discriminator, we use a classical neural network with sigmoid output activation whose output value ranges from 0 to 1 in accordance to the binary classification task. With a reasonable choice of the number of hidden units, this simple neural network architecture has enough capacity for the discriminative task.

\subsection{Adversarial Differentiable Training}\label{learning}
Generative adversarial training of the quantum circuit is a quantum-classical hybrid algorithm with feedback loops. The zero-sum minimax game [\Eq{minmax}] does not perform well in practice for classical GANs due to the gradient of the generators vanishes when the discriminator confidently rejects the samples of the generator. Therefore, we employ the heuristic non-saturating loss for the generator following the classical GAN literature~\cite{Goodfellow2016c}. The objective functions to be minimized become
\begin{eqnarray}
L_{D_{\boldsymbol{\phi}}} & = & - \E_{x\thicksim\pi}[ \ln D_{\boldsymbol{\phi}}(x)]  - \E_{x \thicksim p_{\boldsymbol{\theta}}}[\ln(1 - D_{\boldsymbol{\phi}}(x))], \label{loss_D} \\
L_{G_{\boldsymbol{\theta}}}  & = &  - \E_{x \thicksim p_{\boldsymbol{\theta}}}[ \ln D_{\boldsymbol{\phi}}(x)] \label{loss_G}.
\end{eqnarray}
In the adversarial training procedure, we minimize them simultaneously using stochastic gradient descent (SGD) which requires gradient information $\nabla_{\boldsymbol{\phi}} L_{D_{\boldsymbol{\phi}}}$ and $\nabla_{\boldsymbol{\theta}} L_{G_{\boldsymbol{\theta}}}$. Since the discriminator is a classical neural network, we can use the backpropagation~\cite{Rumelhart1986} to calculate  $\nabla_{\boldsymbol{\phi}} L_{D_{\boldsymbol{\phi}}}$ efficiently.

Since one only has direct access to the measured bitstrings $x$, but not the output probability $p_{\boldsymbol{\theta}}(x)$ of a quantum device, it is nontrivial to compute the gradient of the generator $\nabla_{\boldsymbol{\theta}} L_{G_{\boldsymbol{\theta}}} = -\sum_{x} \ln D_{\boldsymbol{\phi}}(x)\nabla_{\boldsymbol{\theta}}p_{\boldsymbol{\theta}}(x)$ on an actual quantum device.
Fortunately, for the parametrized quantum circuit considered in Sec.~\ref{model}, the gradient of the output probability is~\cite{Liu2018}
\begin{eqnarray}
\label{prob_gradient}
\nabla_{\theta_{\alpha}^{l}}p_{\boldsymbol{\theta}}(x) = \frac{1}{2} \left(p_{\boldsymbol{\theta}^{+}}(x) - p_{\boldsymbol{\theta}^{-}}(x)\right),
\end{eqnarray}
where $\boldsymbol{\theta}^{\pm}= \boldsymbol{\theta} \pm \frac{\pi}{2}\boldsymbol{e}_{\alpha}^{l}$, with $\boldsymbol{e}_{\alpha}^{l}$ the $(\alpha, l)$-th unit vector in parameter space.
Thus, even without direct access to the probability $p_{\boldsymbol{\theta}}$, one still obtains an unbiased gradient estimator for the quantum circuit generator
\begin{eqnarray}
\label{score_function_estimator_real}
\nabla_{\boldsymbol{\theta}} L_{G_{\boldsymbol{\theta}}} = \frac{1}{2}\E_{x\thicksim p_{\boldsymbol{\theta}^-}} [\ln D_{\boldsymbol{\phi}}(x)] - \frac{1}{2}\E_{x\thicksim p_{\boldsymbol{\theta}^+}} [\ln D_{\boldsymbol{\phi}}(x)].
\end{eqnarray}
Notice that this is different from the approximated finite difference estimators of the gradient. In practice, we just need to tune the quantum circuit's parameters to $\boldsymbol{\theta}^{\pm}$ and then measure a batch of samples to estimate the gradient. Here, the estimator is similar to the score function gradient estimator in deep learning literature, which applies broadly to explicit density models with continuous and discrete variables~\cite{Schulman2015GradientEU}. However, due to the unique feature of the quantum circuit [\Eq{prob_gradient}], one can obtain an unbiased gradient estimator even for an implicit probability distribution.

\begin{algorithm}[H]
	\begin{algorithmic}	
      \caption{Learning of the Generative Adversarial Quantum Circuit}
        \Require A quantum circuit with $N$ qubits of depth $d$ for generative modeling. A classical neural network for binary classification. The learning rate $\eta$.
        \Ensure  Parameters of the quantum circuit generator $G_{\boldsymbol{\theta}}$ and classical neural network discriminator $D_{\boldsymbol{\phi}}$.
	    \While{$\boldsymbol{\theta}, \boldsymbol{\phi}$ have not converged}
         \State {\color{blue}{\texttt{\#train D}}}
	     \State A batch of samples from the dataset $x\thicksim \pi$
         \State A batch of samples from the circuit $x\thicksim p_{\boldsymbol{\theta}}= |\langle x | \psi_{\boldsymbol{\theta}} \rangle|^{2}$
         \State Compute $\nabla_{\boldsymbol{\phi}} L_{D_{\boldsymbol{\phi}}}$ via backpropagation  
         \State $\boldsymbol{\phi}= \boldsymbol{\phi} -\eta\nabla_{\boldsymbol{\phi}} L_{D_{\boldsymbol{\phi}}}$
        \State {\color{blue}{\texttt{\#train G}}}
        \State Tune circuit parameters to $\boldsymbol{\theta}^{+}$, sample $x\thicksim p_{\boldsymbol{\theta^+}}= |\langle x | \psi_{\boldsymbol{\theta^+}} \rangle|^{2}$
        \State Tune circuit parameters to $\boldsymbol{\theta}^{-}$, sample $x\thicksim p_{\boldsymbol{\theta^-}}= |\langle x | \psi_{\boldsymbol{\theta^-}} \rangle|^{2}$
        \State Compute $\nabla_{\boldsymbol{\theta}} L_{G_{\boldsymbol{\theta}}}$ using  \Eq{score_function_estimator_real}
        \State $\boldsymbol{\theta}=\boldsymbol{\theta} -\eta \nabla_{\boldsymbol{\theta}} L_{G_{\boldsymbol{\theta}}}$
         \EndWhile
		\label{alg:trainning}
	\end{algorithmic}
\end{algorithm}
The Algorithm listing \ref{alg:trainning} summarizes the adversarial training of the quantum circuit for generative modeling.


\subsection{Quantum Inference}
The trained quantum circuit represents classical probability as a quantum pure state $|\psi\rangle=\sum_x \sqrt{p(x)}|x\rangle$, denoted as q-sample in Ref.~\cite{Low2014}. We have omitted the $\boldsymbol{\theta}$ subscript for conciseness. The quantum circuit can generate new samples according to the probability upon projective measurement. One can also use the quantum circuit for inference, i.e. infer the value of unobserved qubits conditioned on partial observations on some qubits. To make the discussion concrete, consider a division of the qubits into two sets $x=q\cup e$. Inference amounts to sample the query $q$ according to the conditional probability $p(q|e)$ given the evidence $e$.

\begin{figure}[tb]
    \centerline{\includegraphics[width=0.45\textwidth]{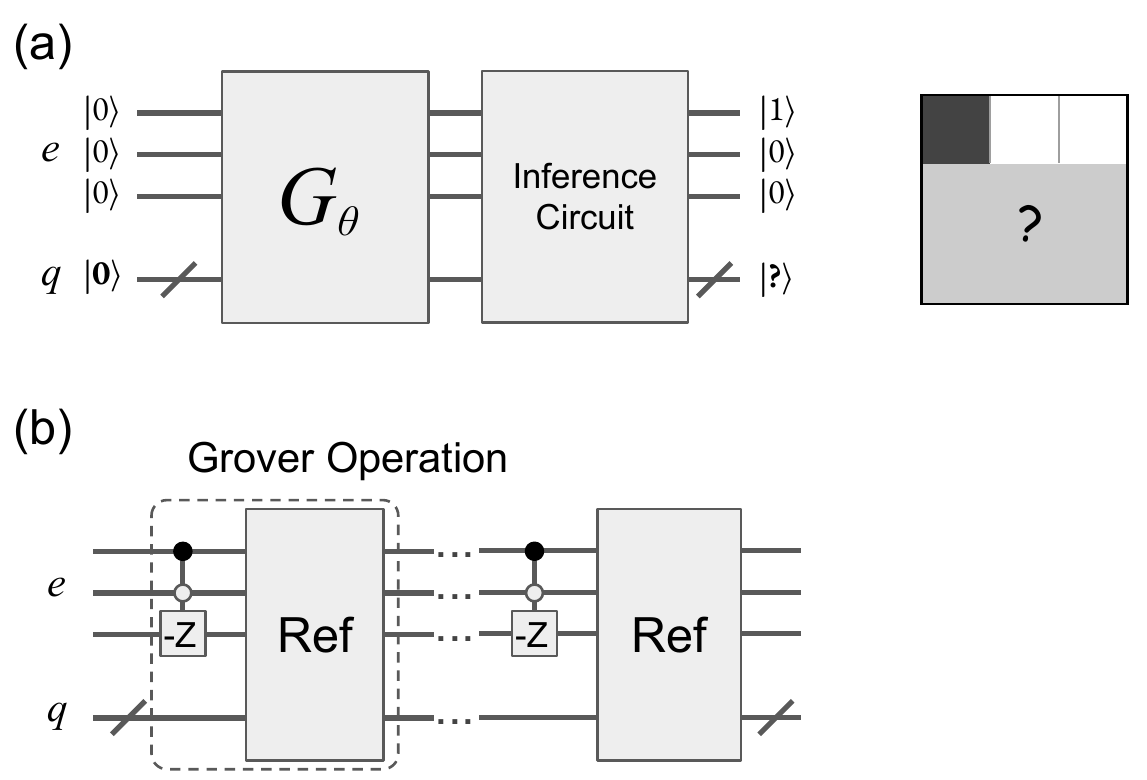}}
    \caption{(a) Illustration of the quantum inference task. Given the value of evidence qubits, e.g., the first three lines, one attempts to predict the value of remaining qubits according to the probability of the generative quantum circuit. (b) Inference circuit enhance the probability collapsing into the subspace in accordance to the evidence via amplitude amplification. For the control gate, filled dots represent control qubits and circled dots represent inverse control qubits. "Ref" is the Householder reflection operation.}
\label{inference-illu}
\end{figure}

A straightforward sampling approach would be jointly measuring all the qubits and post selecting those samples agree with the given evidence. However, encoding the classical probability in a quantum state allows quadratic quantum speed up in inference compared to this naive rejective sampling approach. This was originally described for quantum inference of classical Bayesian networks using quantum circuits~\cite{Low2014}. Given a trained QCBM, we rewrite the output state as $|\psi\rangle = \sqrt{p(e)}|q, e\rangle+\sqrt{1-p(e)}|\overline{q, e}\rangle$, $|q, e\rangle$ is the target space where the basis is in accordance to the evidence, and $|\overline{q, e}\rangle$ is the space orthogonal to the target space. Here we have absorbed the relative sign into the basis $|q, e\rangle$ and $|\overline{q, e}\rangle$.
The quantum inference on a Born machine with amplitude amplification consists of the following two steps. See Fig.~\ref{inference-illu} for an illustration of the procedure.

\begin{enumerate}

    \item Apply oracle circuit on the evidence qubits. The oracle is a multi-controlled Z (or -Z) gate that flip the sign of wave function in the target space.
    For example in Fig. \ref{inference-illu} (b), this oracle acts on the first 3 qubits and flips the sign of wave function whenever the configuration these qubits is "100".
   The resulting state is $|\chi\rangle=-\sqrt{p(e)}|q,e\rangle + \sqrt{1-p(e)}|\overline{q,e}\rangle$

    \item Perform Householder reflection ${\rm Ref} = 2|\psi\rangle\langle\psi|-1$ on $|\chi\rangle$.
    The construction of Householder reflection utilizes the learned circuit ${\rm Ref}=U_{\boldsymbol{\theta}} (2|z\rangle\langle z|-1)U_{\boldsymbol{\theta}}^\dagger$.
   The resulting state after this operation is
\begin{align}
    \begin{split}
        {\rm Ref}|\chi\rangle= &\left [ 3-4p(e) \right]\sqrt{p(e)}|q,e\rangle\\
        +& \left[1-4p(e)\right]\sqrt{1-p(e)}|\overline{q,e}\rangle
    \end{split}\label{eq-grover}
\end{align}

\end{enumerate}
From \Eq{eq-grover}, we see that as long as the marginal probability before iteration $p(e)<1/2$, the probability of collapsing into the target space will increase after the Grover operation. We thus repeat the two steps until we obtain the desired evidence with a probability close to 1. The number of Grover operations is inverse proportional to the square root of the original marginal probability of evidence.
Meanwhile, the desired sector $|q, e\rangle$ is untouched in this inference process. Thus, whenever projective measurement on the evidence qubits obtains $e$, the remaining qubits will follow the correct conditional probability $p(q|e)$.

\section{Numerical Experiments}\label{sec-result}
We implement the generative adversarial learning and inference of quantum circuit using a quantum circuit simulator to demonstrate its feasibility. We also discuss practical issues of the hardware implementation of the generative adversarial quantum circuits.

\subsection{\BAS Dataset}
We train the generative adversarial quantum circuit on the discrete Bars-and-Stripes dataset~\cite{MacKay2002, Han2017, Liu2018,Benedetti2018}, which is a synthetic image dataset contains either bars or stripes on a two-dimensional grid. For $m\times n$ pixels, there are only $2^m + 2^n -2$ valid \BAS patterns among $N_{basis} = 2^{m\times n}$  possible images. This defines the target probability distribution $\pi(x)$, which is a constant for valid images and zero for invalid images. 
The number of qubits need in the QCBM is the same as the number of pixels $m\times n$. We train the QCBM with BAS datasets of size $2\times2$, $2\times3$ and $3\times3$.

\begin{figure}[tb]
\centerline{\includegraphics[width=0.45\textwidth]{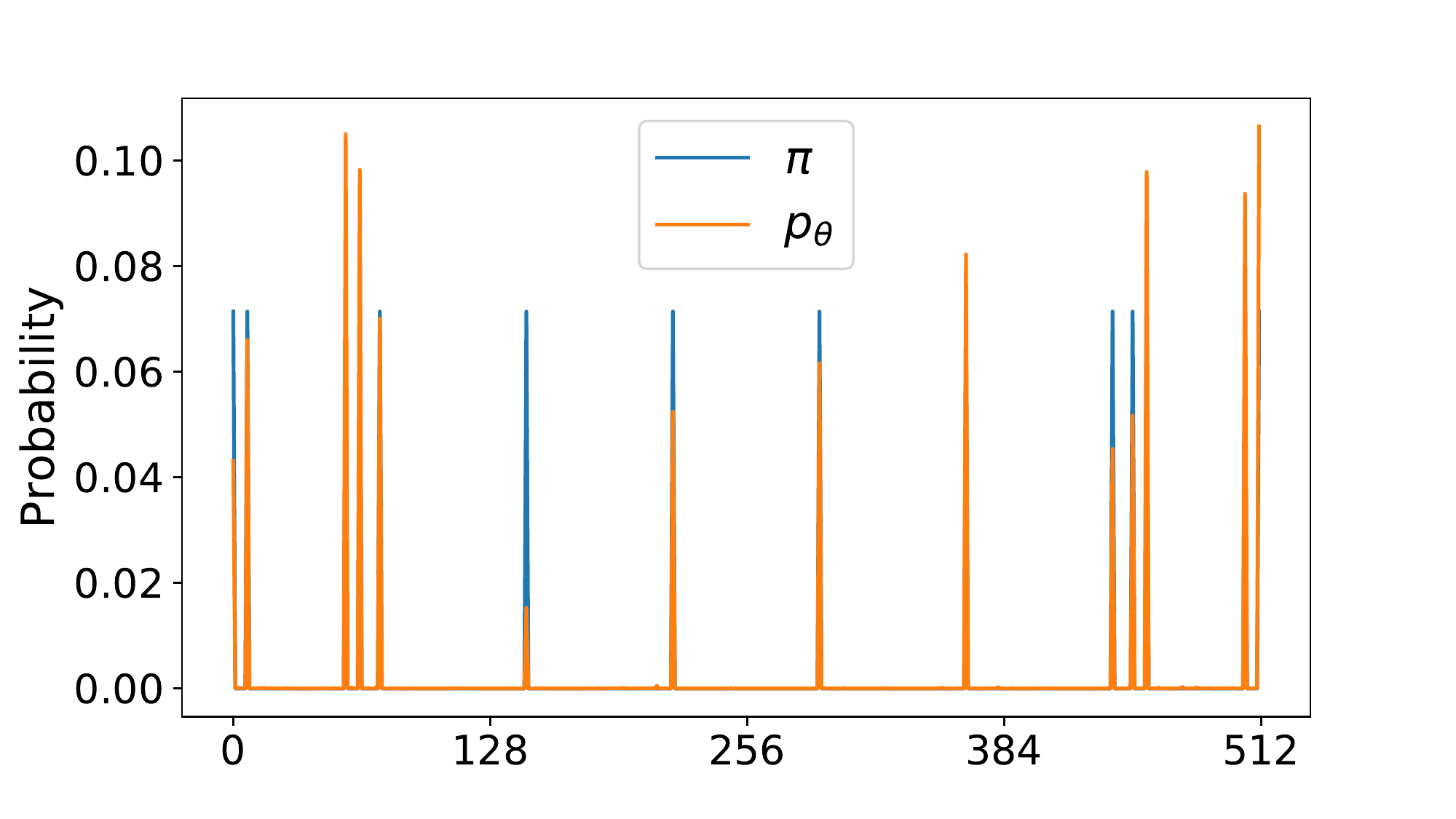}}
\caption{The probability distribution of $3\times3$ \BAS dataset. $\pi$ is the probability distribution of the target data, while $p_{\boldsymbol{\theta}}$ is the probability distribution learned by the quantum circuit after 10000 iterations.  The quantum circuit's depth is $d=28$ and and batch size is $b=512$}
\label{distribution}
\end{figure}

\subsection{Adversarial Training}
For the generator G, we use the quantum circuit described in Section~\ref{model}.
To design the connectivity of the entangler layers, we consider arranging the $N$ qubits in $m\times n$ square grid with periodic boundary condition and connect the nearest neighbor qubits.
The connection pairs are \{(0, 1), (1, 2), (2, 0), (3, 4), (4, 5), (5, 3), (6, 7), (7, 8), (8, 6), (0, 3), (3, 6), (6, 0), (1, 4), (4, 7), (7, 1), (2, 5), (5, 8), (8, 2)\} on $3\times 3$ square grid. The control bit of each pair is the first element. We note alternative ways of designing and compiling the circuit architecture is possible~ \cite{Liu2018}.
For the discriminator D, we use a fully connect neural network with two hidden layers. There are 64 leaky rectified linear units in each hidden layer. 

\begin{figure}[tb]
\centerline{\includegraphics[width=0.45\textwidth]{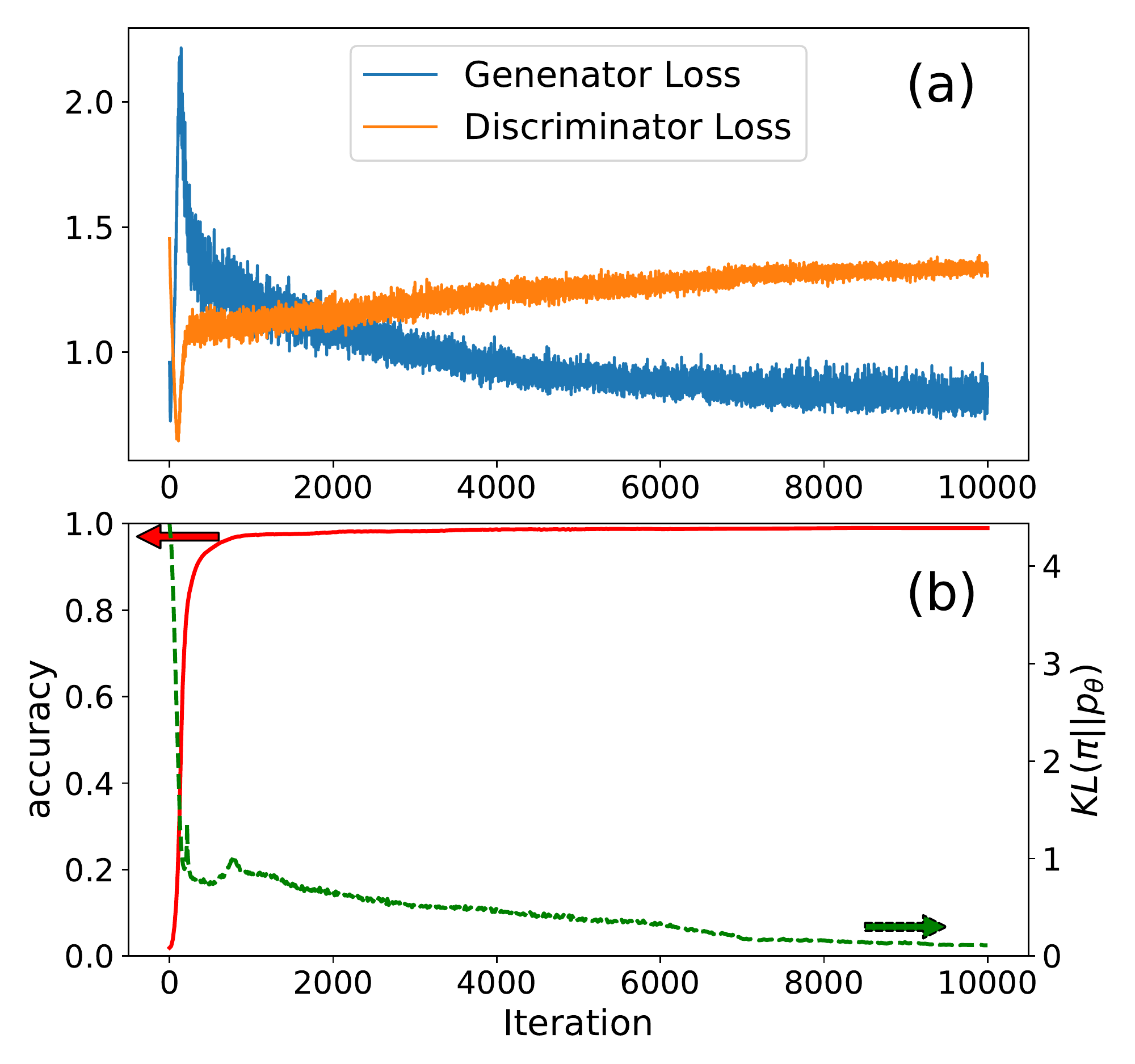}}
    \caption{The result of learning a $3\times3$ \BAS dataset with quantum circuit depth $d=28$ and batch size $b=512$. (a) The discriminator and generator loss with respect to iterations. (b) The accuracy (red) measures the ratio of generated samples which belong to the \BAS pattern. The Kullback-Leibler divergence (green) $\mathbb{KL}(\pi || p_{\boldsymbol{\theta}} )$ between the target probability distribution $\pi$ and the generator distribution $p_{\boldsymbol{\theta}}$. }
\label{accuracy_dynamic}
\end{figure}

For the optimization, we employ the Adam optimizer with the learning rate $10^{-4}$ for both the discriminator and the generator. We update the discriminator and the generator parameters alternatively at each iteration. The remaining hyper-parameters are the batch size $b$ and the depth of the quantum circuit $d$.

In Figure~\ref{distribution}, we show the learned probability of the the $3\times3$ dataset with quantum circuit depth $d=28$ and batch size $b=512$. The output probability of the circuit is suppressed almost to zero for invalid configurations, which means that with the help of the discriminator, the generator learns to produce valid configurations. However, the model probability $p_{\boldsymbol{\theta}}(x)$ does not perfectly match $\pi(x)$, some peaks are smaller than expected. This phenomenon, known as the mode collapse~\cite{Goodfellow2016c, Hong2018GANReview, Lin2017PacGAN} is one of the major challenges in the training of classical GANs.

To characterize the training dynamics, Figure~\ref{accuracy_dynamic} (a) shows the loss function of the generator and the discriminator [Eqs.~(\ref{loss_D}, \ref{loss_G})]. Both of them saturate after 10000 iterations. And the D loss converges to a value around 1.34, which is about twice of the G loss. This could be understood
since near the equilibrium $D(x) \approx 1/2$ for any of the inputs and one obtains $L_{D_{\boldsymbol{\phi}}} \approx 2L_{G_{\boldsymbol{\theta}}} \approx 2\ln 2$.
Figure~\ref{accuracy_dynamic} (b) shows the Kullback-Leibler (KL) divergence between $\pi$ and $p_{\boldsymbol{\theta}}$, $\mathbb{KL}(\pi || p_{\boldsymbol{\theta}})=\sum_x \pi(x)  \ln [{\pi}(x)/{ p_{\boldsymbol{\theta}}}(x)]$, which is a measure of the dissimilarity between the two distributions. The KL-divergence decreases continuously towards zero with training. The KL-divergence is not directly measurable in an actual experiment since it involves the output probability of the circuit. We thus define the accuracy as the percentage of the valid samples belong to the \BAS pattern in the samples generated by the quantum circuit. Figure~\ref{accuracy_dynamic} (b) shows that the accuracy increases continuously towards one in the training. 

\begin{figure}[tb]
\centerline{\includegraphics[width=0.45\textwidth]{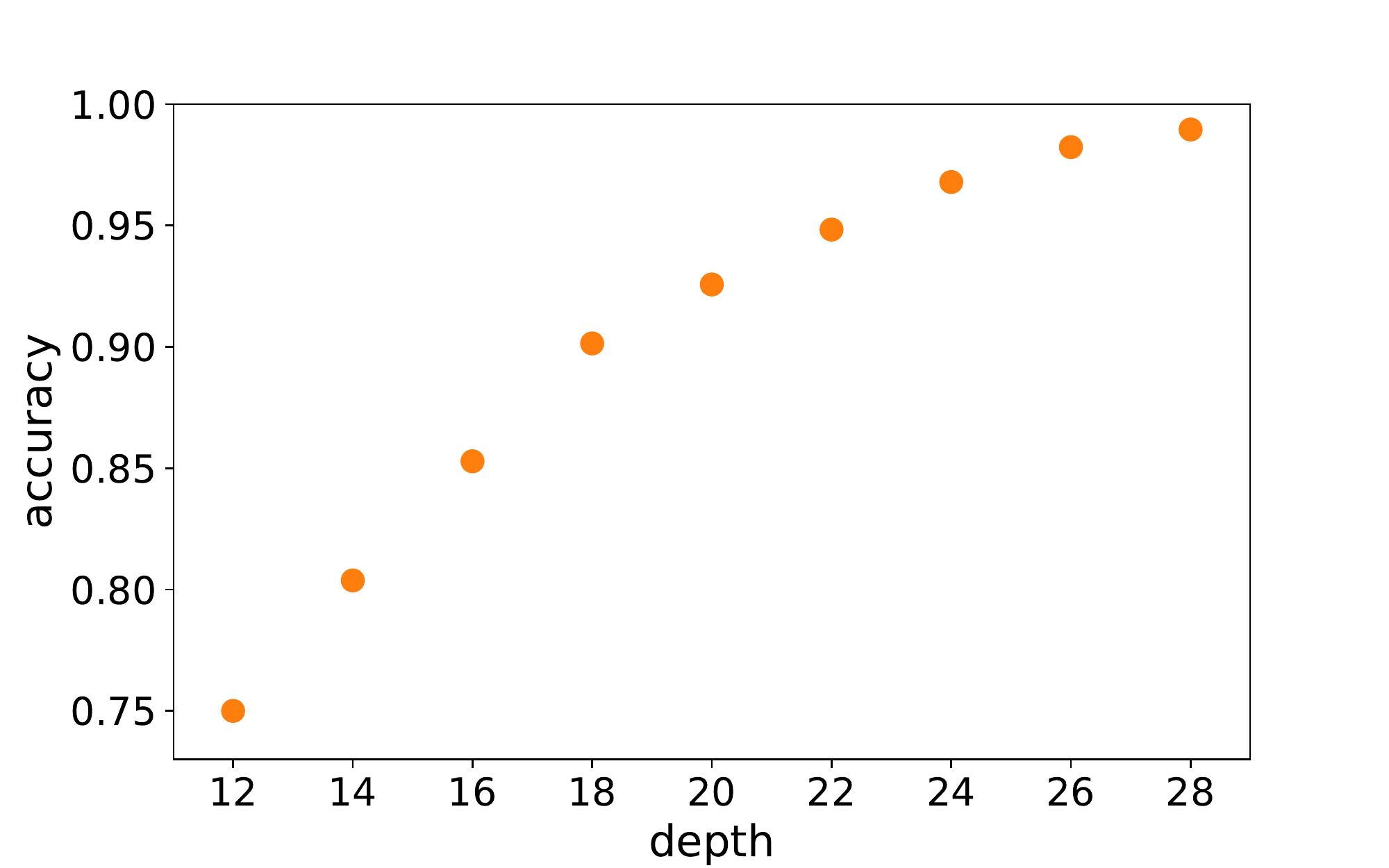}}
\caption{The accuracy versus quantum circuit depth $d$ for the $3\times3$ \BAS dataset.}
\label{accuracy_vs_d}
\end{figure}

Next, we train generative adversarial quantum circuits of different depth $d$ on a dataset of $3\times3$ \BAS and monitor the final accuracy. The results summarized in Fig.~\ref{accuracy_vs_d} show that the accuracy increases with the circuit depth, which suggests that one can memorize the training dataset with a sufficiently deep quantum circuit.

\begin{table}[tb]
\centering
\addtolength{\tabcolsep}{12pt}
\caption{Accuracy of the generative adversarial quantum circuit of different image size $m\times n$, and $N$ is the number of qubits. $N_{basis}$ is the dimension of the Hilbert space. $d$ is the depth of the quantum circuit. $N_{\theta}$ is the number of learnable parameters in the quantum circuit. For $2\times2$, $2\times3$ and $3\times3$ datasets, we choose batch size $b=64$, $b=128$ and $b=512$ respectively.}
\label{tab:hyperparameter}
\begin{tabular}{@{}llllll@{}}
\toprule
$m \times n $ & $N$  & $N_{basis}$  & $d$       & $N_{\theta}$  & accuracy    \\ \midrule
$2\times2$    & 4             & 16           & 2       & 28            &  99.97\%     \\
$2\times3$    & 6             & 64           & 5       & 96            &  99.71\%    \\
$3\times3$    & 9             & 512          & 28      & 765           & 98.96\%     \\ \bottomrule
\end{tabular}
\end{table}

To test the expressibility of the circuit depth with respect to the size of the input, we train the generative adversarial quantum circuit on $2\times2$, $2\times3$ and $3\times3$ \BAS datasets. We aim to push the accuracy of the valid samples to $100\%$ for these small scale datasets. The results summarized in Table~\ref{tab:hyperparameter} implies that the number of parameters in the quantum circuit is of the same order of the number $N_{basis}$, which is exponential with the image size. As a result, one may need even deeper quantum circuits to fully overfit image dataset of larger size.

For practical applications on future intermediate size circuits, the goal is not to memorize the training dataset, but to obtain the generalization ability on unseen data. Therefore, the results observed for the small dataset is not necessarily a problem. On the contrary, shallow quantum circuits provide means of regularization to prevent overfitting. In any case, exploring tensor networks inspirations~\cite{2017arXiv171107500K, Huggins2018}, circuit structure learning~\cite{Liu2018} and Bayesian networks~\cite{Low2014, Du2018} can be helpful for putting correct inductive biases into the quantum circuit architecture.

\subsection{Quantum Inference}
\begin{figure}[tb]
\centerline{\includegraphics[width=0.45\textwidth]{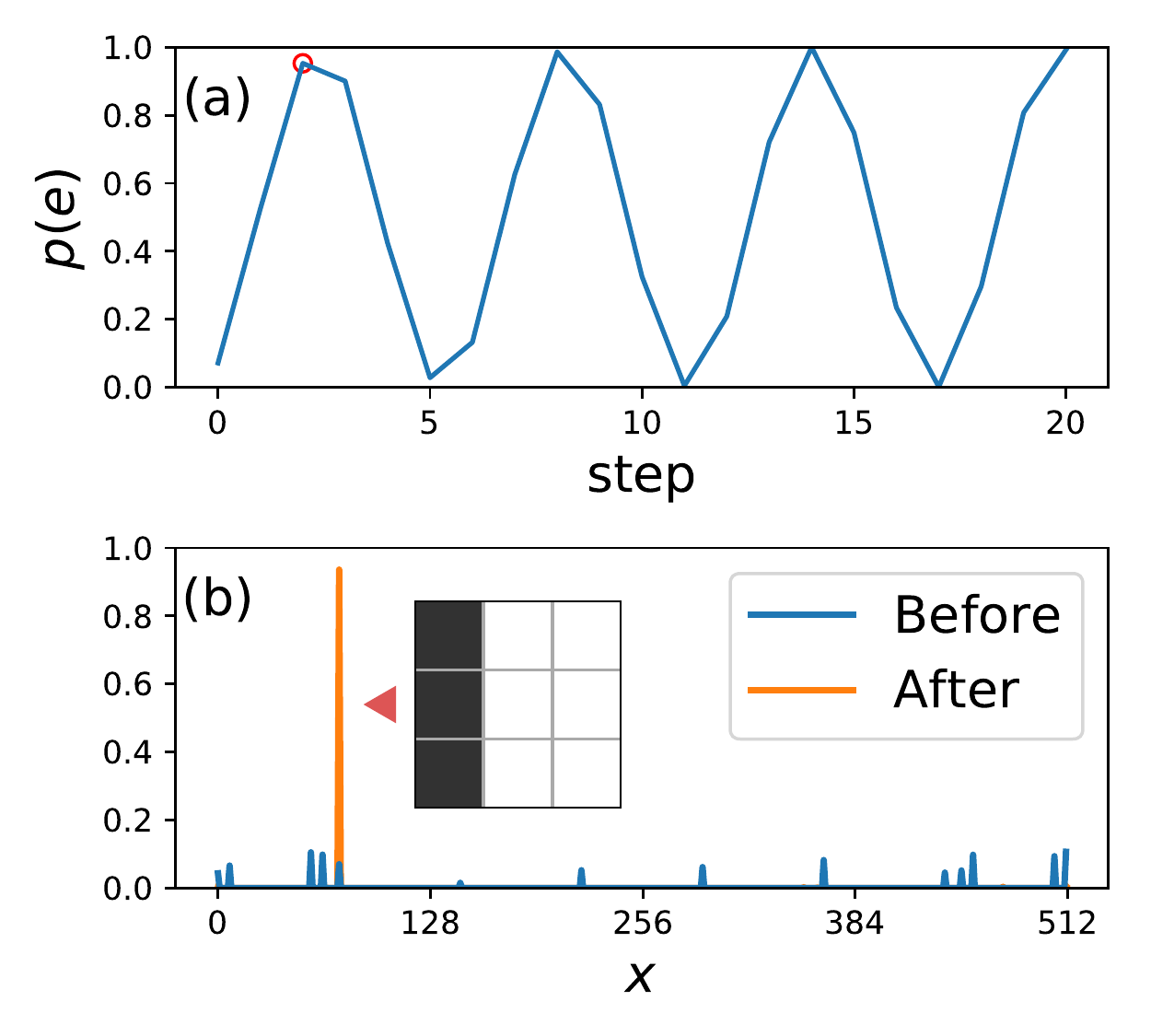}}
    \caption{
    (a) The marginal probability of evidence as a function of the number of Grover operations for $3\times 3$ \BAS dataset and circuit depth 28.
    (b) The blue curve is the output probability before inference. The orange curve is the probability after two Grover operations. The annotated peak is the desired configuration.
    }
\label{3by3}
\end{figure}

Once we trained a quantum circuit which captures the probability distribution of the \BAS dataset in Fig.~\ref{distribution}, we can use it to perform image inpainting. For example, given an incomplete configuration
$$
x = \left[
\begin{matrix}
1 & 0 &0 \\
\cdot & \cdot  &\cdot  \\
\cdot & \cdot  &\cdot
\end{matrix}
\right],
$$
where $\cdot$ indicate missing values, we perform quantum inference to restore the missing values following the procedure of Fig.~\ref{inference-illu}.
As shown in Fig.~\ref{3by3} (a), after two Grover operations, the marginal probability of the evidence has reached a maximum $95.3\%$ (red circled dot).
Fig.~\ref{3by3} (b) shows that the probability on the target space is enlarged by a factor of 13.4 compared to direct measuring on the quantum state without amplitude amplification. Next, suppose the measurement gives the desired evidence, with probability $98.3\%$ (the ratio between the height of orange peak and 95.3\%) one will obtain the correct configuration corresponding to the position of the orange peak
$$
x = \left[
\begin{matrix}
1 & 0 &0 \\
1 & 0 &0 \\
1 & 0 &0
\end{matrix}
\right].
$$
The additional failure probability is due to the imperfection of training, which leave small but non-vanishing amplitudes on the invalid configurations with desired evidence. These configurations are all amplified by the same factor as the orange peak, which ensures the conditional probability $p(q|e) = {p(q, e)}/{p(e)}$.

For the actual experiment realization, most of the inference circuit is straightforward to implement except for the multi-controlled gates shown in Fig.~\ref{inference-illu}. It can nevertheless be decomposed into a polynomial number of basic control gates. Another practical issue is how to determine the exact number of Grover steps in the actual experiment since the probability will oscillate as one applies the Grover iteration. This can be solved by using exponential searching~\cite{Brassard2002, Low2014}, where the amortized running complexity remains $\mathcal{O}(1/\sqrt{p(e)})$.

\section{Discussions}\label{sec-discussion}
Generative adversarial learning of quantum circuits is a fresh approach to quantum machine learning~\cite{Biamonte2017} in the age of noisy intermediate-scale quantum technologies~\cite{Preskill2018}. Our numerical experiments on the \BAS dataset provide baseline results for the training and inference on small scale quantum circuits. A natural next step would be implementing these algorithms on actual quantum devices.
Moreover, one also needs to explore other shallow quantum circuit architectures with optimal expressive capacity~\cite{2017arXiv171107500K, Huggins2018, Liu2018, Du2018} and alternative training objective functions for better performance.

Finally, we considered a fixed product state $|z\rangle = |0 \rangle^{\otimes N}$ as the initial state of the generative quantum circuit. Thus, all the uncertainty is due to the inherently probabilistic nature of the quantum mechanics. In future, it is also worth exploring the case where the input state follows a classical prior probability distribution $p(z)$, and the model probability is $p_{\boldsymbol{\theta}}(x) = \sum_z p(z) | \langle x| U_{\boldsymbol{\theta}}|z \rangle|^2 $. With the input basis state $z$ acting as latent variables the model can capture even richer class of probabilities. And having a quantum latent space can support a larger class of unsupervised learning tasks. 


\emph{Note--} During preparation of this manuscript, we note an arXiv preprint on related topic~\cite{Situ2018}.

\section{Acknowledgment}
Learning and inference of the generative adversarial quantum circuits are implemented using \texttt{Yao.jl}~\cite{Julia}, a flexible, extensible, efficient framework for quantum algorithm design written in Julia language. We thank Xiu-Zhe (Roger) Luo for help with \texttt{Yao.jl}. JGL and LW are supported by the Ministry of Science and Technology of China under the Grant No. 2016YFA0300603, the National Natural Science Foundation of China under the Grant No. 11774398, and the research program of the Chinese Academy of Sciences under Grant No. XDPB0803. JFZ and JPH are supported by the Ministry of Science and Technology of China 973 program (No. 2015CB921300, No.~2017YFA0303100), National Science Foundation of China (Grant No. NSFC-1190020, 11534014, 11334012), and the Strategic Priority Research Program of CAS (Grant No.XDB07000000). YFW is supported by the Undergraduate Innovation Program of CAS (Grant No.Y7CY031D31).

\bibliographystyle{apsrev4-1}
\bibliography{qcgan}

\end{document}